\documentclass[pra,showpacs,twocolumn]{revtex4}
\usepackage{graphicx}
\usepackage{dcolumn}
\usepackage{bm}
\usepackage{amssymb}
\usepackage{amsmath}
\usepackage{epsfig}
\usepackage{times}

\input{tcilatex}

\begin{document}
\title{Indistinguishability of independent single photons}
\author{F. W. Sun\footnote{%
fs2293@columbia.edu}}
\author{C. W. Wong}
\affiliation{Optical Nanostructures Laboratory, Columbia University, New York, NY 10027. }
\date{\today }

\begin{abstract}
The indistinguishability of independent single photons is presented by
decomposing the single photon pulse into the mixed state of different
transform limited pulses. The entanglement between single photons and outer
environment or other photons induces the distribution of the center
frequencies of those transform limited pulses and makes photons
distinguishable. Only the single photons with the same transform limited
form are indistinguishable. In details, the indistinguishability of single
photons from the solid-state quantum emitter and spontaneous parametric down
conversion is examined with two-photon Hong-Ou-Mandel interferometer.
Moreover, experimental methods to enhance the indistinguishability are
discussed, where the usage of spectral filter is highlighted.
\end{abstract}

\pacs{42.50.Ar, 42.25.Hz, 03.65.Yz}
\maketitle

\section{Introduction}

Linear optical quantum computation \cite{Kok07RMP} is based on the
interference between different photons \cite{HOM}, in which the
indistinguishability of photons is a fundamental and critical requirement.
Any distinguishability will reduce the visibility of interference and the
fidelity of quantum computation protocol \cite{Rohde05PRA}. It will also
directly affect the other applications with photon interference, such as
quantum key distribution \cite{Gisin02RMP} and high precision quantum phase
measurement \cite{Sun08EPL}. Moreover, photon indistinguishability is
fundamental to stimulated emission \cite{Lamas-Linares01NAT, Sun07PRL} and
has been applied in quantum cloning \cite{Lamas-Linares02SCI,Fasel02PRL} and
entanglement measure \cite{Sun07PRA2}. Based on the spontaneous parametric
down conversion (SPDC), the indistinguishability in the multiphoton
interference has been intensely examined recently in experiment \cite%
{Ou99PRL,Tsujino04PRL,Eisenberg06PRL,Xiang06PRL,Liu07EPL} and theory \cite%
{Ou99PRA,OuPRAs,Sun07PRA,Sun08PRA}. However, the kernel is the
indistinguishability of independent single photons. In SPDC, independent
single photons are heralded by detecting the twinning photons, with several
experiments focusing on their indistinguishability and interference \cite%
{Riedmatten03PRA,Kaltenbaek06PRL,Mosley08PRL}. In the solid-state quantum
emitters, single photons have been remarkably examined \cite%
{Martini96PRL,Kim99NAT,Michler00SCI,Santori01PRL,Kuhn02PRL,Santori02NAT},
where, in addition to photon statistics and quantum efficiencies,
indistinguishability is another important character of the single photon
source \cite{Santori02NAT,Mosley08PRL}.

Generally, the distinguishability of the single photons comes from the
entanglement with extrinsic system, such as photons, phonons or outer
environment. Theoretically, the single photons can be described as the mixed
state by tracing out the entangling parts. In SPDC, the property of
entanglement can be achieved through the analysis of the phase matching
condition \cite{U'Ren07PRA,Mosley08PRL}. However, it is more complicated in
the solid-state quantum emitters. Many kinds of physical processes introduce
the entanglement between the environment and the emitted single photons. For
example, in the single quantum dot, the interaction with phonon results in
short dephasing time and gives rises to a very broad spectrum of the photons
\cite{Carmichael}. This spectrum broadening will make photons
distinguishable. Since it is hard to measure all the physical information of
each photon source, the direct analysis of the photon state with
interference is highly desired.

In this paper, we will give a brief description of single photons to show
the indistinguishability. In the frequency degree of freedom (DOF), the
whole photon state is a mixed state of transform limited pulses with
different center frequencies. For two independent single photons, there is
no entanglement between them. The indistinguishability describes the nature
of identicality of the transform limited pulses. To aid in the analysis, we
regarded the bandwidth of distribution of these center frequencies as the
extrinsic width, which comes form the entanglement between with extrinsic
system. The transformed limited pulse are pure state and its width is the
intrinsic width. For the same single photon source, single photons have the
same extrinsic width and the same intrinsic width. The total spectrum
bandwidth is the combination of the intrinsic width and the extrinsic width.
Generally, when the extrinsic width is much larger than the intrinsic width,
the single photons are totally distinguishable. Only when the extrinsic
width is zero, the single photon pulse is transform limited and
indistinguishable. In either Lorentzian or Gaussian distributions of the
spectrum, the photon indistinguishability is the ratio of intrinsic width to
total bandwidth. In experiment, the distinguishability can be measured with
Hong-Ou-Mandel (HOM) interferometer \cite{HOM}, where the visibility shows
the indistinguishability. In the main section, we will examine the
indistinguishability of single photons from solid quantum emitters and SPDC
after a general description of the single photon state is given. In the
discussion section, experimental methods to enhance the indistinguishability
are presented, where the effect of spectral filter is highlighted.

\section{Description and indistinguishability of single photons}

We begin the description of single photons from the transform limited pulse,
which is a pure quantum state,%
\begin{equation}
\left\vert \omega \right\rangle =\int_{-\infty }^{+\infty }d\upsilon
g_{\omega }(\upsilon )a^{\dag }(\upsilon )\left\vert \text{vac}\right\rangle
\text{,}
\end{equation}%
where\ $a^{\dag }$ ($a$) is the single photon creation (annihilation)
operator. $|g_{\omega }(\upsilon )|^{2}$ is the spectrum of the transform
limited pulse with center frequency $\omega $ and width $\Delta _{g}$
(intrinsic width). We will discuss the independent single photons from the
same source and assume the same $\Delta _{g}$, since the interactions
between the single photons and outer environment or other photons are highly
similar during the generation. Correspondingly, the transform limited pulse
has the duration of $T_{TL}=1/\Delta _{g}$. Also, $g_{\omega }(\upsilon )$
satisfies the normalization condition $\int_{-\infty }^{+\infty }d\upsilon
|g_{\omega }(\upsilon )|^{2}=1$. The indistinguishability of two independent
transform limited photon pulse is $K_{ij}^{TL}=\left\vert \left\langle
\omega _{i}|\omega _{j}\right\rangle \right\vert ^{2}$. Roughly, the two
photons are totally distinguishable when $\left\vert \omega _{i}-\omega
_{j}\right\vert \gg \Delta _{g}$ and indistinguishable for $\omega
_{i}=\omega _{j}$.

\begin{figure}[t]
\includegraphics[width=7.5cm]{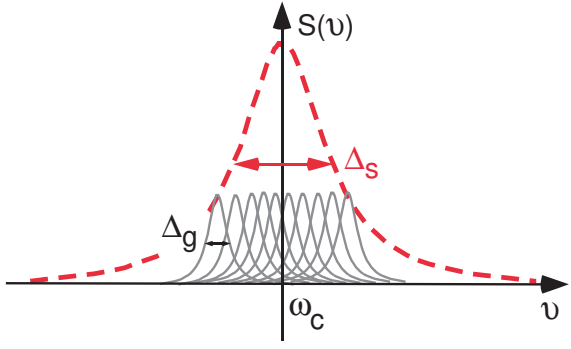}
\caption{(color online) Illustration of total single photon pulse (red
dashed curve, width $\Delta _{s}$) composed of transform limited pulses
(grey bold curves, width $\Delta _{g}$) with different center frequencies.
When $\Delta _{S}=\Delta _{g}$, the single photon pulse is transform limited
and indistinguishable.$\,$}
\label{Pulse}
\end{figure}

Since the single photons may be entangled with extrinsic system, the center
frequencies have the distribution $f(\omega )$ [$\int_{-\infty }^{+\infty
}d\omega f(\omega )=1$] with width $\Delta _{f}$ (extrinsic width). Then,
the whole state is written as
\begin{equation}
\rho =\int_{-\infty }^{+\infty }d\omega f(\omega )\left\vert \omega
\right\rangle \left\langle \omega \right\vert \text{.}  \label{TS}
\end{equation}%
The total spectrum $S(\upsilon )=\int_{-\infty }^{+\infty }d\omega f(\omega
)\left\vert g_{\omega }(\upsilon )\right\vert ^{2}$ is broadened to $\Delta
_{S}\geq $ $\Delta _{g}$ because of the distribution $f(\omega )$. However,
the lifetime of the single photon pulse is same with those transform limited
pulses, that is $T_{\rho }=T_{TL}$. Fig. \ref{Pulse} illustrates that the
total single photon pulse is composed of different transform limited pulses.
Only when $\Delta _{S}=\Delta _{g}=1/T_{\rho }$ is satisfied, the single
photon pulse becomes transform limited.

Formally, the indistinguishability of two independent single photons can be
described as
\begin{equation}
K=tr(\rho \otimes \rho )=\iint\nolimits_{-\infty }^{+\infty }d\omega
_{i}d\omega _{j}f(\omega _{i})f(\omega _{j})\left\vert \left\langle \omega
_{i}|\omega _{j}\right\rangle \right\vert ^{2}\text{.}  \label{Dis}
\end{equation}%
If and only if $\rho $ is the pure state, $K=1$. That is with $\Delta _{f}=0$
and $\Delta _{S}=$ $\Delta _{g}$, the single photon states are
indistinguishable. On the other hand, when $\Delta _{S}\gg $ $\Delta _{g}$, $%
K\rightarrow 0$, the single photon states are distinguishable. From this
view, two photons may be distinguishable even when they have the same
description. The indistinguishability describes the nature of identicality
of the pure state. Eq. (\ref{Dis}) also describes the purity of the state $%
\rho $. Generally, a mixed state comes from an entangled system. In
principle, distinguishable information of the state may be obtained by
measuring the entangling part. Therefore, the mixed state always has some
distinguishability and the purity of the state is a good scale to evaluate
the indistinguishability. Moreover, this definition of indistinguishability
is highly supported by the experiment. In experiment, the indistinguishable
photons will present photon bunching effect and the value of the
indistinguishability has the simple relationship with the interference
visibility. Based on the single photons from solid-state quantum emitters
and SPDC, we will now give detailed discussions on their
indistinguishability.

\subsection{Indistinguishability of single photons from single solid quantum
emitter}

Here we focus on the single photons from single quantum dot. The single
photon from two-level quantum dot spontaneous emission has the Lorentzian
distribution,
\begin{equation}
g_{\omega }(\upsilon )=\frac{1}{\sqrt{\pi }}\frac{\sqrt{\Gamma /2}}{%
(\upsilon -\omega )+i\Gamma /2}\text{,}
\end{equation}%
where $\Gamma /2$ is the intrinsic width and describes the rate of
spontaneous emission \cite{Scully}. Correspondingly, the lifetime is $%
T_{1}=1/\Gamma $. In addition to the intrinsic linewidth, the spectrum
broadening mainly comes from the dephasing process. Also, the spectral
diffusion of single quantum dot gives much more broader spectrum \cite%
{Neuhauser00PRL}. All these spectrum broadening can be included in the
distribution of $f(\omega )$. For simplicity, we only consider the spectral
broadening from pure dephasing which can also be described as the Lorentzian
function,%
\begin{equation}
f(\omega )=\frac{1}{\pi }\frac{\Gamma ^{\prime }}{(\omega -\omega
_{c})^{2}+\Gamma ^{\prime 2}}\text{,}
\end{equation}%
where $\omega _{c}$ is the center frequency of the distribution $f(\omega )$%
. The extrinsic width is $\Gamma ^{\prime }=1/T_{2}^{\prime }$, where $%
T_{2}^{\prime }$ is the pure dephasing time. The total state can be
described with Eq. (\ref{TS}). The whole spectrum is
\begin{equation}
S(\upsilon )=\frac{1}{\pi }\frac{\Gamma _{2}}{(\upsilon -\omega
_{c})^{2}+\Gamma _{2}^{2}}\text{,}
\end{equation}%
where $\Delta _{S}^{L}=\Gamma _{2}=1/T_{2}=\Gamma ^{\prime }+\Gamma /2$ is
the total spectral width and the superscript $L$ in $\Delta _{S}^{L}$
denotes Lorentzian distribution. In the time domain, we get $%
1/T_{2}=1/2T_{1}+1/T_{2}^{\prime }$. When $\Gamma ^{\prime }=0$, $\Gamma
_{2}=\Gamma /2=1/2T_{1}$, the single photons are transform limited.

The indistinguishability of the two transform limited pulses centered at $%
\omega _{i}$ and $\omega _{j}$ is
\begin{equation}
K_{ij}^{TL}=\frac{\Gamma ^{2}}{(\omega _{i}-\omega _{j})^{2}+\Gamma ^{2}}%
\text{,}
\end{equation}%
while the indistinguishability of the two single photons with Eq. (\ref{Dis}%
) is%
\begin{equation}
K_{L}=\frac{\Gamma }{2\Gamma _{2}}\text{.}
\end{equation}%
When $\Gamma ^{\prime }=0$, $K_{L}=1$ and the single photon are the
transform limited and indistinguishable.

\begin{figure}[t]
\includegraphics[width=8.5cm]{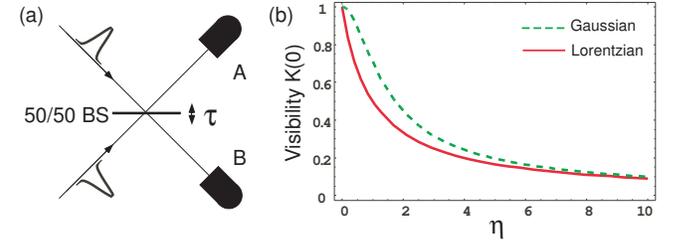}
\caption{(color online)(a) Illustration of two-photon Hong-Ou-Mandel
interference. $\protect\tau $ is the interval between arrival time of the
two input photons. (b) Two-photon Hong-Ou-Mandel interference visibility
with different ratios ($\protect\eta $) of external width to intrinsical
width. $\protect\eta =\protect\eta _{L}=2\Gamma ^{\prime }/\Gamma $ for
Lorentzian distribution and $\protect\eta =\protect\eta _{G}=\protect\sigma %
_{f}/\protect\sigma _{g}$ for Gaussian distribution. The visibility
approaches $1/\protect\eta $ when $\protect\eta $ is much larger than $1$.}
\label{HOM}
\end{figure}

Experimentally, the HOM interferometer is usually used to measure the
indistinguishability of two single photons, as shown in Fig.\ref{HOM}(a).
Two single photons are injected into the two input ports of a 50/50
beamsplitter separately. The two-photon coalescence probability $C_{AB}$ of
output ports$\ A$ and $B$ is null when two photons are indistinguishable and
arrive at the beamsplitter simultaneously. Any distinguishability will
induce nonzero two photon coalescence probability and reduce the
interference visibility. In order to obtain the coalescence probability $%
C_{AB}(\tau )$ with the interval $\tau $ between the arrival times of two
photons, we first calculate probability of two photons exiting in the same
output port $C_{AA}(\tau )$, which shows photon bunching when the two
photons are indistinguishable \cite{Ou99PRA,HBT}. Therefore,
\begin{widetext}

\begin{eqnarray}
C_{AA}(\tau ) &=&\frac{1}{8}\left\langle E^{(-)}(t)E^{(-)}(t+\tau
)E^{(+)}(t+\tau )E^{(+)}(t)\right\rangle   \notag \\
&=&\frac{1}{8}\iint\nolimits_{-\infty }^{+\infty }d\omega
_{i}d\omega _{j}f(\omega _{i})f(\omega _{j})\int_{-\infty }^{+\infty
}dt \left\langle \omega _{i}\right\vert \left\langle \omega
_{j}\right\vert E^{(-)}(t)E^{(-)}(t+\tau )E^{(+)}(t+\tau
)E^{(+)}(t)\left\vert \omega
_{j}\right\rangle \left\vert \omega _{i}\right\rangle   \notag \\
&=&\frac{1}{4}\iint\nolimits_{-\infty }^{+\infty }d\omega
_{i}d\omega _{j}f(\omega _{i})f(\omega _{j})\int_{-\infty }^{+\infty
}dt \lbrack \left\langle \omega _{j}\right\vert
E^{(-)}(t)E^{(+)}(t)\left\vert \omega _{j}\right\rangle \left\langle
\omega _{i}\right\vert E^{(-)}(t+\tau
)E^{(+)}(t+\tau )\left\vert \omega _{i}\right\rangle   \notag \\
&&+\left\langle \omega _{j}\right\vert E^{(-)}(t)E^{(+)}(t+\tau )\left\vert
\omega _{i}\right\rangle \left\langle \omega _{i}\right\vert E^{(-)}(t+\tau
)E^{(+)}(t)\left\vert \omega _{j}\right\rangle ]  \notag \\
&=&\frac{1}{4}[1+K(\tau )]
\end{eqnarray}%
\end{widetext}where $E^{(+)}(t)=\int_{-\infty }^{+\infty }d\omega a(\omega
)e^{-i\omega t}/\sqrt{2\pi }$ is the detection operator. The coefficient $%
1/8 $ comes from photon loss of the beamsplitter ($1/4$) for two photons and
the normalization coefficient of two permutations of two photons detecting
by two detectors ($1/2$). In the practical experiment, the detection
duration is much larger than the photon pulse lifetime and the integral time
is extended to $(-\infty ,+\infty )$. In the above equation, $K(\tau )$ is
the indistinguishability of two photons with time interval $\tau $

\begin{eqnarray}
K(\tau ) &=&\iint\nolimits_{-\infty }^{+\infty }d\omega _{i}d\omega
_{j}f(\omega _{i})f(\omega _{j})\int_{-\infty }^{+\infty }dt  \notag \\
&&\times \left\vert \left\langle \omega _{i}\right\vert
E^{(-)}(t)E^{(+)}(t+\tau )\left\vert \omega _{j}\right\rangle \right\vert
^{2}\text{.}  \label{K}
\end{eqnarray}%
For the Lorentzian distribution, the indistinguishability is
\begin{equation}
K_{L}(\tau )=\frac{\Gamma }{2\Gamma _{2}}e^{-\Gamma |\tau |}\text{,}
\end{equation}%
and the two-photon probability is
\begin{equation}
C_{AA}(\tau )=\frac{1}{4}(1+\frac{\Gamma }{2\Gamma _{2}}e^{-\Gamma |\tau |})%
\text{.}
\end{equation}%
The excess probability of $K(\tau )$ is the signature of the photon
indistinguishability. It is the result of photon bunching from the
permutation symmetry of bosonic particles \cite{Sun07PRA,Sun08PRA}.

Because of the symmetry of the beamsplitter, the probability of two photons
together in the output port $B$ is same with $C_{AA}(\tau )$. Therefore, the
two-photon coincidence probability $C_{AB}(\tau )$ based on the energy
conservation law is
\begin{eqnarray}
C_{AB}(\tau ) &=&1-2C_{AA}(\tau )  \notag \\
&=&[1-K(\tau )]/2 \\
&=&\frac{1}{2}(1-\frac{\Gamma }{2\Gamma _{2}}e^{-\Gamma |\tau |})\text{.}
\label{HOM-L}
\end{eqnarray}%
$C_{AB}(\tau )$ shows the typical HOM dip with the $1/e$ width of pulse
lifetime, $1/\Gamma =T_{1}$, as shown in the experimental report \cite%
{Santori02NAT}. The visibility shows the indistinguishability of $K_{L}$,
which is illustrated in Fig. \ref{HOM}(b) with different ratios of extrinsic
width to intrinsic width, $\eta =\Delta _{f}/\Delta _{g}$. For Lorentzian
distribution, $\eta=\eta _{L}=2\Gamma ^{\prime }/\Gamma $. When extrinsic
width is much larger than the intrinsic width, $\eta _{L}\gg 1$, the
indistinguishability is approaching to $1/\eta _{L}$.

\subsection{Indistinguishability of single photons from SPDC}

In the SPDC, the distinguishability of single photons is induced by the
entanglement between the twinning photon. In order to obtain the information
of the heralded single photon, the entangling parts need to be traced out in
theory. From SPDC, the two-photon state can be written as \cite%
{Ou97QSO,U'Ren07PRA}%
\begin{equation}
\left\vert S,I\right\rangle =\iint\nolimits_{-\infty }^{+\infty }d\omega
_{S}d\omega _{I}\Phi (\omega _{S},\omega _{I})a_{S}^{\dag }(\omega
_{S})a_{I}^{\dag }(\omega _{I})\left\vert \text{vac}\right\rangle \text{,}
\label{SPDC}
\end{equation}%
where $\Phi (\omega _{S},\omega _{I})=P(\omega _{S}+\omega _{I})H(\omega
_{S},\omega _{I})$ is the two-photon wave function, which contains the
information of the pump beam spectrum $P(\omega _{S}+\omega _{I})$ and the
phase matching condition $H(\omega _{S},\omega _{I})$ in the nonlinear
crystal. We assume the pump beam is transform limited and the spectrum $%
|P(\omega _{S}+\omega _{I})|^{2}$ is Gaussian distribution with width $%
\sigma _{g}$. By making the detection of the idle photon ($I$) with a single
frequency of $\Omega _{I}$, the signal photon ($S$) has the transform
limited single photon state from Eq.(\ref{SPDC}),

\begin{equation}
\left\vert S\right\rangle _{\Omega _{2}}=\int_{-\infty }^{+\infty }d\omega
_{S}P(\omega _{S}+\Omega _{I})H(\omega _{S},\Omega _{I})a_{S}^{\dag }(\omega
_{S})\left\vert \text{vac}\right\rangle \text{.}
\end{equation}%
Under the normal phase matching condition for thin nonlinear crystal, the
bandwidth of $H(\omega _{S},\Omega _{I})$ is much larger than the pump width
\cite{Grice97PRA,Kim05OL}. Therefore, the $H(\omega _{S},\Omega _{I})$ is
slowly varying function and can be taken outside of the integral. In this
case, the transform limited single photon pulse has the same shape and width
of the pump beam, which can be described with $g_{\omega }(\upsilon )=$ $%
e^{-(\upsilon -\omega )^{2}/4\sigma _{g}^{2}}/\sqrt[4]{2\pi \sigma _{g}^{2}}$%
.

Since the actual detection of the idle photon is the sum of the above
detections of different frequency $\Omega _{2}$, the center frequency of the
transform limited single photon pulse has the distribution of $f(\omega )$.
Without loss of generality, we assume that $f(\omega )=e^{-(\omega -\omega
_{c})^{2}/2\sigma _{f}^{2}}/\sqrt{2\pi \sigma _{f}^{2}}$ \cite%
{Grice97PRA,Kim05OL}. Therefore, the heralded single photon can be formally
described in Eq. (\ref{TS}) with intrinsic width $\sigma _{g}$ and extrinsic
width $\sigma _{f}$. The total spectrum is also Gaussian distribution with
the width
\begin{equation}
\sigma =\sqrt{\sigma _{g}^{2}+\sigma _{f}^{2}}\text{.}
\end{equation}%
Moreover, the indistinguishability of two photons with interval $\tau $\ is
calculated,
\begin{equation}
K_{G}(\tau )=\frac{\sigma _{g}}{\sigma }e^{-\tau ^{2}\sigma _{g}^{2}}\text{.}
\label{Dis-Gau}
\end{equation}%
The two-photon coalescence probability $C_{AB}(\tau )$ for the HOM
interference is%
\begin{equation}
C_{AB}(\tau )=\frac{1}{2}(1-\frac{\sigma _{g}}{\sigma }e^{-\tau ^{2}\sigma
_{g}^{2}})  \label{HOM-G}
\end{equation}%
with the visibility of $K_{G}(0)=\sigma _{g}/\sigma $, which is also shown
in Fig. \ref{HOM}(b) with different ratios of $\eta _{G}=\sigma _{f}/\sigma
_{g}$. Moreover, the indistinguishability approaches to $1/\eta _{G}$ with
large extrinsic width. If the two-photon wave function can be factorized,
i.e. $\Phi (\omega _{S},\omega _{I})=\Phi (\omega _{S})\Phi (\omega _{I})$,
the single photons is transform limited \cite{Ou97QSO,U'Ren07PRA,Mosley08PRL}%
. In this case, $\sigma _{f}=0$ and the single photons are indistinguishable.

From the results of HOM interference, Eq. (\ref{HOM-L}) and Eq. (\ref{HOM-G}%
), the width of indistinguishability, or the two-photon fourth-order
coherence, only depends on the intrinsic width, or the lifetime of the
transform limited pulse. However, the total single photon spectral width
determines the width of the single photon second-order coherence time.

\section{Discussion}

\subsection{The definition and the experimental enhancement of
indistinguishability}

From the single photon state, the indistinguishability is described in Eq. (%
\ref{Dis}), which is the purity of the state if the single photons are
generated in the same source. For the single photons from the different
source, the indistinguishability has the description of $K_{ij}=tr(\rho
_{i}\otimes \rho _{j})$. At the same time, from the multi-mode theory, the
indistinguishability is described as $\mathcal{E}/\mathcal{A}$, where $%
\mathcal{E}$($\mathcal{A}$) is the excess (accidental) two-photon
probability \cite{Ou99PRA}. In Refs. \cite{Sun07PRA,Sun08PRA}, the
indistinguishability is derived from the coefficients of Schmidt
decomposition. All of these definitions are equivalent.

It needs to be emphasized that the extrinsic spectral width comes from the
entanglement with extrinsic system. Only this extrinsic spectral width will
bring the distinguishability. In the above discussion, we assumed that all
other DOFs of the single photon have the same states and no entanglement
with the frequency DOF. Actually, the entanglement between the frequency DOF
and other inner DOF of the same photon may induce the mixed spectrum
description. However, for the same entanglement, the mixed spectrum will not
induce the distinguishability when all the DOFs are included, since the
entangled state can be described as a linear superposition form for the
single photon in a higher dimensional space.

Practically, in order to enhance the indistinguishability, different methods
are needed to narrow the extrinsic spectral width or broaden the intrinsic
spectral width. For the quantum emitters, low temperature is needed to
reduce the interaction with phonons. In this case, the dephasing time is
extended \cite{Borri01PRL} and the extrinsic spectral broadening is
controlled. Moreover, the interaction with optical cavity mode will decrease
the lifetime of the spontaneous emission through Purcell effect \cite%
{Purcell46PR}. Therefore, the intrinsic width is broadened and the
indistinguishability is enhanced \cite{Santori02NAT}. In SPDC, particular
design on the phase matching condition helps to generate indistinguishable
single photons \cite{U'Ren07PRA,Mosley08PRL}. However, the usage of spectral
filter is the most feasible method to enhance the indistinguishability,
especially in the experiment on SPDC.

\subsection{The effect of spectral filter}

In experiment, the narrow spectral filter is widely used to enhance the
indistinguishability and interference visibility. Theoretically, the
detection operator after the spectral filter can be described as%
\begin{equation}
E^{(+)}(t)=\frac{1}{\sqrt{2\pi }}\int_{-\infty }^{+\infty }d\omega F(\omega
)a(\omega )e^{-i\omega t}\text{,}
\end{equation}%
where $|F(\omega )|^{2}$ is the spectral transmissivity of the filter. Here
we assume the Gaussian distribution of $|F(\omega )|^{2}=e^{-(\omega -\omega
_{c})^{2}/2\sigma _{F}^{2}}$, centered same at $\omega _{c}$ with width $%
\sigma _{F}$.

With the spectral filter, the spectrum of the single photons is also
Gaussian distribution and its width narrows to
\begin{equation}
\sigma ^{\prime }=\frac{\sigma _{F}\sqrt{\sigma _{g}^{2}+\sigma _{f}^{2}}}{%
\sqrt{\sigma _{g}^{2}+\sigma _{f}^{2}+\sigma _{F}^{2}}}\text{.}
\end{equation}%
At the same time, the filter narrows the intrinsic width,
\begin{equation}
\sigma _{g}^{\prime }=\frac{\sigma _{F}\sigma _{g}}{\sqrt{\sigma
_{g}^{2}+\sigma _{F}^{2}}}\text{.}  \label{In-F}
\end{equation}%
Using Eq.(\ref{Dis-Gau}), the indistinguishability is
\begin{equation}
K_{G}^{\prime }=\frac{\sigma _{g}\sqrt{\sigma _{g}^{2}+\sigma
_{f}^{2}+\sigma _{F}^{2}}}{\sqrt{\sigma _{g}^{2}+\sigma _{F}^{2}}\sqrt{%
\sigma _{g}^{2}+\sigma _{f}^{2}}}\text{,}
\end{equation}%
for $\tau =0$.

More rigorously, the effect of the spectral filter extended from Eq.(\ref{K}%
) is described as:%
\begin{eqnarray}
K_{G}^{\prime } &=&\iint\nolimits_{-\infty }^{+\infty }d\omega _{i}d\omega
_{j}f(\omega _{i})f(\omega _{j})\int_{-\infty }^{+\infty }dt  \notag \\
&&\times \left\vert \left\langle \omega _{j}\right\vert
E^{(-)}(t)E^{(+)}(t)\left\vert \omega _{i}\right\rangle \right\vert
^{2}/C^{2}  \label{KF} \\
&=&\iint\nolimits_{-\infty }^{+\infty }d\omega _{i}d\omega _{j}f(\omega
_{i})f(\omega _{j})  \notag \\
&&\times \left\vert \int d\upsilon F(\upsilon )g_{\omega _{i}}(\upsilon
)F^{\ast }(\upsilon )g_{\omega _{j}}^{\ast }(\upsilon )\right\vert ^{2}/C^{2}
\notag \\
&=&\frac{\sigma _{g}\sqrt{\sigma _{g}^{2}+\sigma _{f}^{2}+\sigma _{F}^{2}}}{%
\sqrt{\sigma _{g}^{2}+\sigma _{F}^{2}}\sqrt{\sigma _{g}^{2}+\sigma _{f}^{2}}}%
\text{,}
\end{eqnarray}%
where $C$ is the probability to detect the single-photon after the filter%
\begin{eqnarray}
C &=&\int_{-\infty }^{+\infty }d\omega f(\omega )\int_{-\infty }^{+\infty
}dt\left\langle \omega \right\vert E^{(-)}(t)E^{(+)}(t)\left\vert \omega
\right\rangle   \label{C} \\
&=&\frac{\sigma _{F}}{\sqrt{\sigma _{g}^{2}+\sigma _{f}^{2}+\sigma _{F}^{2}}}%
\text{.}
\end{eqnarray}%
Certainly there is photon loss for $C<1$ when using the filter to enhance
the indistinguishability.
\begin{figure}[t]
\includegraphics[width=8.5cm]{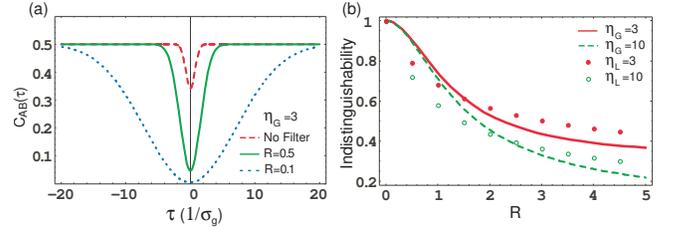}
\caption{(color online) Indistinguishability with different widths of
filters. (a) Two photons coalescence probability $C_{AB}(\protect\tau )$ of
the HOM interference without filter (red dashed curve) and with filters
(green solid curve for $R=0.5$ and blue dotted curve for $R=0.1$). Here we
set $\protect\eta _{G}=3$ for all three cases. The two curves with filters
are normalized to the maximal probability of $1/2$ for total distinguishable
cases ($\protect\tau \gg \protect\sigma _{g}^{\prime }$). (b) The red solid
and the green dashed curves show the indistinguishability for $\protect\eta %
_{G}=3$ and $\protect\eta _{G}=10$ with Gaussian filters, respectively. The
red solid ($\protect\eta _{L}=3$) and green open ($\protect\eta _{L}=10$)
circles are for the corresponding results of Lorentzian filter on Lorentzian
spectrum distribution. In this case, $R=2\Gamma _{F}/\Gamma $, where $\Gamma
_{F}$ is the Lorenzian filter width.}
\label{Dis-Filter}
\end{figure}

Fig. \ref{Dis-Filter} shows the effect on the indistinguishability with
different ratios of spectral filter width to intrinsic width, $R=\sigma
_{F}/\sigma _{g}$. In Fig. \ref{Dis-Filter}(a), the width of HOM dip is
broadened to $1/\sigma _{g}^{\prime }$, since the intrinsic width is
narrowed by the spectral filter in Eq.(\ref{In-F}). Fig. \ref{Dis-Filter}(b)
shows the indistinguishability with $\eta _{G}=3$ (red solid) and $\eta
_{G}=10$ (green dashed). In comparison, the results of the Lorentzian filter
on Lorentzian spectrum distribution are also shown in Fig. \ref{Dis-Filter}%
(b) with red solid ($\eta _{L}=3$) and green open ($\eta _{L}=10$) circles.
These results are numerically calculated with Eq.(\ref{KF}) and Eq.(\ref{C}%
). Clearly, the indistinguishability is approaching to $1$ when the filter
width is closing to $0$. For $\eta \gg 1$, the value of the
indistinguishability shows the same result as in Fig. \ref{HOM}(b), where
the extrinsic width is replaced by the filter width.

In SPDC, for the pump pulse duration of $110$fs (full width at half
maximum), the indistinguishability with a full width at half maximum $3$nm
filter is about $0.94$ for $\eta \gtrsim 3$ \cite{Kim05OL}. It is little
higher than the experimental results in \cite{Xiang06PRL,Sun08EPL} because
there may be entanglement in other degrees of freedom between the twin
photons besides the frequency entanglement \cite{Sun07PRA}. Here, we used
the condition that the single photon intrinsic width $\sigma _{g}$ is same
with the pump beam width for thin nonlinear crystal.

\subsection{Independent photons from many quantum emitters}

In some cases, there is more than one independent photon from many quantum
emitters. The total state is
\begin{equation}
\rho _{N}=\prod_{i=1}^{N}\rho _{k}\text{.}
\end{equation}%
where $\rho _{k}=(C\left\vert \text{vac}\right\rangle \left\langle \text{vac}%
\right\vert +\int_{-\infty }^{+\infty }d\omega f_{k}(\omega )\left\vert
\omega \right\rangle \left\langle \omega \right\vert )$ is for the
independent single photon with $C=1-\int_{-\infty }^{+\infty }d\omega
f_{k}(\omega )$. Considering the photon loss in the practical experiment and
quantum efficiency of the quantum emitters, $\int_{-\infty }^{+\infty
}d\omega f_{k}(\omega )<1$. Moreover, $f_{k}(\omega )$ may have different
center wavelengthes. For example, there is size distribution of quantum
dots. In this case, the total spectrum will include the broadening from size
distribution. Therefore, the spectrum is very broad and the photons will be
distinguishable even at the low temperature.

\section{Conclusion}

The description of the single photons state in the spectrum domain is
presented to discuss the indistinguishability. The ratio of extrinsic
spectrum width to intrinsic width governs the indistinguishability. Single
photons are indistinguishable only when they have the same transform limited
forms, while they are highly distinguishable when the extrinsic spectrum
width is much larger than the intrinsic width. Fundamentally, the
indistinguishability of independent photons shows the sameness of part which
can be described with pure state and only the indistinguishable parts can
interfere each other. In experiment, the indistinguishability shows excess
probability of two-photon coincident detection in Hanbury-Brown-Twiss
interferometer \cite{HBT} or less probability in HOM interferometer.
Moreover, the indistinguishability can be experimentally enhanced with the
narrow spectral filter or by controlling the generation condition.

\begin{acknowledgments}
F.W.S. thanks Z. Y. Ou for helpful discussion. This work is funded by DARPA,
NSF Contract No. ECCS 0747787, and the New York State Office of Science,
Technology and Academic Research.
\end{acknowledgments}

\end{document}